\documentstyle [aps,twocolumn]{revtex}
\begin{document}
\title{Two-dimensional Fermi Gas Revisited}
%%%%%%%%%%%%%%%%%%%%%%%%%%%%%%%%%%%%%%%%%%%%%%%   
\author{Philip W. Anderson}
\address{Joseph Henry Laboratories of Physics\\
Princeton University, Princeton, New Jersey 08544}
%%%%%%%%%%%%%%%%%%%%%%%%%%%%%%%%%%%%%%%%%%%%%%%
\maketitle
\begin{abstract}
A number of authors have taken issue with the demonstration that the
2D Fermion gas with short-range repulsive interactions (and, of
course, including spin) cannot be consistently treated as a
renormalised quasiparticle system.  This paper shows that the
arguments given in some of these papers are invalid or irrelevant.
\end{abstract}
%%%%%%%%%%%%%%%%%%%%%%%%%%%%%%%%%%%%%%%%%%%%%%%
A number of papers have addressed the two-dimensional Fermion gas
from the viewpoint of "Fermi liquid theory" in the sense of
perturbation theory based on renormalised single-particle
excitations.  The first of these was by P. Bloom in 1975 \cite{bloom_75},
the
most recent by Disertori and Rivasseau in 2000 \cite{disertori_00}, based
on earlier
work by Trubowitz et al. \cite{feldman_96},  using the quantum
renormalisation group methods
pioneered by Anderson, Yuval and Hamann \cite{anderson_70}.  I have
argued, on the 
other hand, that there is strong evidence  indicating fundamental 
difficulties with the theory.  The point is made here
that (a) the restriction on the lower cutoff in energy accepted by
refs. \cite{disertori_00} and \cite{feldman_96} excludes specifically the
regions 
of parameter values where the
effects calculated in my work \cite{anderson_92} occur so that 
refs. \cite{disertori_00} and \cite{feldman_96} are not relevant
to the correctness of \cite{anderson_92};  and (b) that the
calculations in reference \cite{bloom_75} (the only attempt at a
rigorously
controlled calculation, to my knowledge, that might have settled the
matter if correct) suffer from an incorrect solution of the
fundamental integral equation.

Some authors \cite{engelbrecht_90}, \cite{yokoyama_97} have accepted my
calculation of 
the interaction vertex and tried to reconcile it with quasiparticle
theory; I also
(c) give a new argument demonstrating the failure of this approach.

The method of \cite{anderson_92} in  fact follows the reasoning of Bloom
in \cite{bloom_75}, who adapted to two dimensions the scheme developed
independently by Galitskii and by Huang, Lee and Yang
\cite{galitskii_58}, 
\cite{huang_57} for
hard-core systems in three dimensions.  This is the "pseudopotential
method" which was demonstrated to give an expansion in powers of the
density, hence is a rigorously controlled approximation.  The basis
of the expansion is the idea that encounters of more than two
particles within each others' range of interaction, a, occupy a
volume of configuration space which is smaller by the factor
$na^D \approx (k_F a)^D$
than binary encounters. In diagrammatic     
terms, diagrams with extra hole lines are smaller by this factor than
those with a single hole line and can be neglected.

Some criticisms have proposed (e.g.\cite{yokoyama_97}) that the
particle-particle 
ladder diagrams responsible for the effects in \cite{anderson_92} could
be cancelled 
by particle-hole loops. The laatter, however, have a different density
dependence
and are irrelevant in the low-density limit.

I should add two remarks about this method.  One is that although the
specific calculations in the original references are carried out for
hard spheres or discs, they are valid a fortiori for any short-range
repulsive potential, as Bloom shows in an appendix to his paper. The
second is that it can be made very plausible that the restriction to
binary interactions is qualitatively generalisable to any density, as
is for instance the case in the fermi liquid theory itself, since the 
Landau functional contains only binary quasiparticle interactions.

The plan of the method of Galitskii and Huang et al (as explained in
the textbook of Fetter and Walecka \cite{fetter_71}, for instance) is
first to
derive the pseudopotential or renormalised vertex
$\Gamma (P ; q,q')$ 
which is the sum of ladder diagrams, that is the solution of the
Bethe-Salpeter equation for the repeated scattering of two particles.
Here P is the total momentum of the incoming particles of momenta p
and p', and q their relative momentum, i. e.
$P = p+p'$ and $q = p - p'$, 
while $q'$ is the relative momentum of the outgoing particles, their
total momentum being P.  The ladder diagrams are the only ones which
survive at this order.  We note, of course, that the important
interactions are only between opposite-spin particles,  since the odd
angular momenta are down by at least a factor $a^2$.  Once having
obtained $\Gamma$, the diagrams for self-energy, chemical potential etc
become simple hartree loops and Bloom, at least, applies well-known
formulae to evaluate these quantities.  (I am not sure whether
the conventional formulas are quite correct for these essentially
on-shell scatterings, but this is not germane to the present
argument.)

Bloom follows Galitskii in solving first for the case of vanishing
density, and then iterating the integral equation for $\Gamma$.  At
vanishing density this first step of course is equivalent to solving
the scattering equation for two free particles of momenta $p$ and $p'$.
Two free particles of opposite spins with momenta $pa$, $p'a<<1$ scatter,
to order $a^2$, in the l=0 channel only, with amplitude
$$
T(q) = -2 exp (i\phi) sin \phi , \eqno(1)
$$
where $\phi$
is the l=0 ``phase shift" for relative momentum and T is the conventional
scattering
matrix. We have chosen the normalisation used by Bloom.  The
interpretation of $\phi$
 as a phase shift is simply a statement of the
boundary condition which the interaction enforces on the wave
function of the two particles at $(r-r') \to 0$,  
and does not directly
make any statement about the imaginary part of any self-energy, or
the like.  For free particles,
$$
\phi_0 \ = \ {2\over t/U + | \ell n qa|}, \eqno(2)
$$
which is essentially the result of the appendix to reference 1.
Here, to make the model more specific, I have used a  tight-binding
Hubbard model of lattice constant a and interaction U, but at low
density this result is model-independent.

It is important to realise that $\phi_0$ in [2] vanishes as $q \to 0$.
This  
is because $T_0$ satisfies an integral equation with a singular kernel,
$$
T_0 (q,q') \ =\ U (q-q') + \int {d^2 p\over (2\pi)^2} \ U (p-q) {T_o
(p,q')\over p^2 - q^2 + 
i\eta} \eqno(3)
$$
whose kernel can  and does diverge, leading to a logarithmic singularity 
in $q$ for the integral term.  From T the vertex
function $\Gamma$ can be derived via an integral representation given in
references (1) and (10), but in fact  its value on the
energy shell is identical to the scattering amplitude T.

When the density is finite the integral equation for the scattering
amplitude T is modified for the effects of the exclusion principle.
The effect of interactions with other particles may be neglected in
our low density approximation, but the statistical "interactions" may
not be ignored because they are effectively of infinite range.  The
integral equation  [3] is modified by including a factor in the
kernel to take account of the exclusion of intermediate states,
$$
N (P, q") \ =\ 1-n \{(P + q")/2\} - n \{(P - q")/2\} \eqno(4)
$$
Here $n$ is the Fermi function, =1 when its argument is less than the
Fermi momentum and 0 otherwise.  As has been shown repeatedly
\cite{anderson_92},
if the momenta p and p' are on the Fermi surface or below it, the
factor N eliminates the singular kernel in the integral equation  and
its solution becomes finite at q=0.  T is given by \cite{bloom_75} with
$$
\phi \ =\ {2\over t/U + |\ell n k_F a|} \eqno(5)
$$
This is the point at which we can dispose of point [1] of the opening
paragraph.  References \cite{disertori_00} and \cite{feldman_96} use
finite 
temperature techniques and are based on
continuation from a weak coupling expansion.  It is necessary (for
reasons I do not quite understand) for them to  restrict the temperature
to
$$
T/T_F \geq  exp (- {t\over U}) \eqno(6)
$$
It seems that with this as a cutoff for the Fermi functions in
\cite{anderson_70},
it would not be possible to distinguish between the phase shifts
\cite{disertori_00} 
and \cite{anderson_92}.  If t/U is sufficiently small that the
logarithmic factor in 
the denominator dominates, $T/T_F$ will be of order unity and there 
will be no sharp fermi surface to exhibit Fermi liquid properties. 
On the other hand, if U is small the cutoff will set in before the 
logarithmic factor dominates, and again one cannot tell whether or 
not the Fermi surface is sharp.

As I said, the reasons for this restriction are not clear. It cannot 
really be to avoid the Kohn-Luttinger divergence in the Cooper 
channel \cite{kohn_65}  because that involves higher l
amplitudes which bring in higher powers of a or of $1/U$.  My suspicion 
is that it may be precisely the divergence of scattering length 
implied by [5] which these works have run up against.

Now let me discuss point (b).  Bloom follows Galitskii in solving 
first for the integral equation equivalent to \cite{feldman_96} for the
vertex 
function $\Gamma_0$,  and then deriving an integral equation for the 
difference between the true $\Gamma$ including the effects of exclusion, 
i e the factor N from \cite{anderson_70}, and $\Gamma_0$. This  procedure
is 
acceptable in 3 dimensions because, although the kernel has a pole, 
the resulting singularity is harmless, amounting to 
$\int {d^3 q\over q^2}$. In 2 dimensions, however, there is a true 
singularity in the kernel, which is precisely the reason for the 
$1/lnq$ behavior of $\Gamma_0$ at q=0. The offending integral equation 
is equation [4] of reference \cite{bloom_75}, which we reproduce here:
$$
\Gamma (p,p' ; P) = \Gamma_0 + \int {d^2k\over (2\pi)^2} \Gamma_0 (p,k ;
P) 
f(P,k) \Gamma (k, p' ; P), \eqno(7)
$$
where 
$$
f(P,k) = 
\left[ {N(P, k)\over \epsilon - k^2 + i\delta N (P,k)} - {1\over 
\epsilon - k^2 + i\delta}\right] 
$$
Naively iterating this equation allows Bloom to obtain a series for 
gamma in powers of $1/\ell nq$,  with each term becoming successively
more 
innocuous.  We note, however, that the kernel in this equation 
subtracts away the same singularity that one has in the zero-density 
case, while the added term is regular for any incoming momenta which 
are on or below the fermi surface.  That is, the iteration is not 
justified because the solution for finite density is qualitatively 
different from that for zero density.  In fact, as we have already 
shown elsewhere, the solution for finite density is perfectly regular 
and does not vanish at $q=0$;  it is given by equations [3[ and [5]. 
This is then the mathematical mistake which led Bloom to conclude 
that the 2 dimensional Fermi gas behaves conventionally.

What is the source of the difficulties of quasiparticle theory that 
this result entails?  That these difficulties are plausible is clear 
on the face of it. The 3-dimensional theory of Huang et al and 
Galitskii brings one straightforwardly to the traditional "scattering 
length" theory, where the scattering length a (not to be confused 
with the interaction range) is defined by
$$
lim (q \to 0) \phi = qa \eqno(8)
$$
The cross-section, for instance, for elastic scattering is 
$\propto a^2$ in 3D. 
Even for zero density, in 2D the scattering length diverges for 
forward scattering;  that is, the concept has no real meaning; but 
Bloom showed that one could nonetheless just get by if it diverged 
only as $1/q \ell nq$.  But in fact it diverges as $1/q$, which means
that the 
scattering length is of the same order as the size of the system: 
like statistical interactions, the range is infinite.  Qualitatively, 
there are always other particles with the same momentum and opposite 
spin which cannot live at the same momentum value because they are 
always within each others' effective scattering length.

Some sets of authors have accepted the demonstration in reference 
\cite{anderson_92} 
of the divergent scattering length, but argued that in one way or 
another one may revive quasiparticle theory nonetheless 
[ \cite{castellani_94} for example].
A paper with Kveshchenko \cite{anderson_95}  gives a bosonisation
technique for 
disproving this, but bosonisation is neither familiar nor totally 
rigorous.  Therefore I give here a rather more straightforward 
demonstration.

What we have demonstrated  above is that there is a finite phase 
shift for {\it elastic}, on-shell scattering in  the $l=0$ channel for 
opposite-spin particles with zero relative momentum.  If such a thing 
were to happen for free particles, it would violate
Levinson's theorem,  but there is no proof that Levinson's theorem
holds in the presence of the exclusions due to the gas of other
particles.

Surprisingly, the appropriate divergences appear in the conventional 
formulas for self-energy as presented in references \cite{bloom_75} 
and \cite{fetter_71}. 
They come from the extension of the scattering amplitude off the 
energy shell, i e from the virtual transitions which cause 
modification of the wave functions rather than real scattering.  It 
is not surprising that they do not contain the exclusion factor 
$N (P, q)$ for this reason. If one simply takes the formula used by Bloom
for the second-order self-energy, his equation [8c],  and inserts the 
correct constant scattering amplitude ,  it is easy to derive that 
the derivative with respect to momentum at fixed energy
 is logarithmically divergent, 
and hence the effective mass is infinite.  The appropriate term in 
Bloom's equation [8c] is
$$
\Sigma_b (p) \ =\ \int\int {d^2k d^2k'\over (2\pi)^4} 
\left\{ T^2 (p, k') N(k) P {1\over k'^2 - q^2} \right\} 
$$
$$
+ {\rm inelastic\ terms} 
$$
Here $q=1/2(p-k)$.  Clearly, setting T equal to the finite constant 
\cite{anderson_92} 
and taking the derivative with respect to p under the integrals leads 
to a logarithmic divergence where the denominator vanishes.
Another indication of this anomaly in $\partial\Sigma / \partial p$ is 
the fact that the effect of the exclusion factor changes radically from 
$P \leq 2k_F$ to $P > 2k_F$, so that there is an infinity in 
$\partial\phi(q) / \partial q$.

Another source of difficulties derives from the optical theorem as 
applied to T.  The existence of a finite scattering amplitude in the 
$l=0$ channel leads to a finite forward scattering, which comes through 
as a delta-function in the effective interaction for $p=p'$.  This 
changes the character of zero sound--it essentially becomes the 
tomographic boson of \cite{anderson_95}.

With a little more effort, the divergence of the renormalisation 
constant Z to zero could be shown.  The divergence of the scattering 
length should however be sufficient to convince the reader that an 
independent particle approximation cannot work, but it is heartening 
that there are in fact obvious perturbative signals to that effect, 
contrary to the many attempts to shore that approximation up: for 
example, ref. \cite{castellani_94}.

Reference \cite{anderson_95} presents a suggestion, tomographic
bozonization, for a 
method which  can deal with these divergences in the case where the 
interaction is not too strong and, particularly, where umklapp is not 
important.  How to go beyond that method, as is necessary to deal 
with the cuprates in particular, is still not obvious.  However, one 
mistake which is often made is to assume that non-Fermi liquid 
behavior and spin-charge separation require strong coupling and 
umklapp effects--these non-fermi liquid properties appear instead to 
be generic in 2 dimensions.

\end{document}